\documentclass[prd,reprint,showpacs,showkeys]{revtex4-1}
\usepackage{amsfonts}
\usepackage{mdframed}
\usepackage{amssymb}
\usepackage{amsmath}
\usepackage{graphicx}
\usepackage[font={footnotesize,it}]{caption}

\begin{document}

\title{Hypocycloidal throat for 2+1-dimensional thin-shell wormholes}
\author{S. Habib Mazharimousavi}
\email{habib.mazhari@emu.edu.tr}
\author{M. Halilsoy}
\email{mustafa.halilsoy@emu.edu.tr}
\affiliation{Department of Physics, Eastern Mediterranean University, Gazima\u{g}usa,
north Cyprus, Mersin 10, Turkey. }
\date{\today }

\begin{abstract}
Recently we have shown that for $2+1-$dimensional thin-shell wormholes a
non-circular throat may lead to a physical wormhole in the sense that the
energy conditions are satisfied. By the same token, herein we consider
angular dependent throat geometry embedded in a $2+1-$dimensional flat
spacetime in polar coordinates. It is shown that a generic, natural example
of throat geometry is provided remarkably by a hypocycloid. That is, two
flat $2+1-$dimensions are glued together along a hypocycloid. The energy
required in each hypocycloid increases with the frequency of the roller
circle inside the large one.
\end{abstract}

\maketitle

\section{Introduction}

Similar to the black holes the wormholes in $2+1-$dimensions \cite{Mann}
also have certain degree of simplicity compared to their $3+1-$dimensional
counterparts \cite{MT}. The absence of gravitational degrees in $2+1-$%
dimensions enforces us to introduce appropriate sources to keep the wormhole
alive against collapse. Instead of general wormholes our concern will be
confined herein to the subject of thin-shell wormholes (TSWs), whose throat
is designed to host the entire source \cite{MV0,MV}. From the outset our
strategy will be to curve the geometry of the throat and find the
corresponding energy-momentum through the Einstein's equations on the
thin-shell \cite{MH1, MH2}. Clearly any distortion / warp at the throat
gives rise to certain source, but as the subject is TSWs the nature of
energy-density becomes of utmost important. Wormholes in general violates
the null-energy condition (NEC) \cite{HV}, which implies also the violation
of the remaining energy conditions. The occurrence of negative pressure
components in $3+1-$dimensions provides alternatives in the sense that
violation of NEC can be accounted by the pressure, leaving the possibility
of an overall positive energy density.

In this paper we choose our throat geometry in $2+1-$dimensional TSW such
that the pressure vanishes, the energy density becomes positive and as a
result all energy conditions are satisfied \cite{MH1}. This is an
advantageous situation in $2+1-$dimensions not encountered in $3+1-$%
dimensional TSWs. Our method is to consider a hypersurface induced in a $%
2+1- $dimensional flat polar coordinates. Upon determining the energy
density it is observed that a natural solution for the underlying geometry
of the throat turns out to be a hypocycloid. Standard cycloid is known to be
the minimum time curve of a falling particle under uniform gravitational
field which is generated by a fixed point on a circle rolling on a straight
line. The hypocycloid on the other hand is generated by a fixed point on a
small circle which rolls inside the circumference of a larger circle. The
warped geometry of such a curve surprisingly generates energy density that
turns out to be positive. This summarizes in brief, the main contribution of
this paper.

In \cite{MH1} we have constructed a $2+1-$dimensional TSW by considering a
flat bulk metric of the form 
\begin{equation}
ds^{2}=-dt^{2}+dr^{2}+r^{2}d\theta ^{2}
\end{equation}%
with a throat located at the hypersurface 
\begin{equation}
F\left( r,\theta \right) =r-a_{0}\left( \theta \right) =0.
\end{equation}%
Using the standard formalism of cut and paste technique (see the Appendix)
it was shown that the line element of the throat is given by%
\begin{equation}
ds_{\Sigma }^{2}=-dt^{2}+\left( a_{0}^{2}+a_{0}^{\prime 2}\right) d\theta
^{2}
\end{equation}%
with the energy momentum tensor on the shell%
\begin{equation}
S_{i}^{j}=\left( 
\begin{array}{cc}
-\sigma _{0} & 0 \\ 
0 & 0%
\end{array}%
\right)
\end{equation}%
in which%
\begin{equation}
\sigma _{0}=\frac{1}{4\pi }\frac{\left( a_{0}^{\prime \prime }-a_{0}-\frac{%
2a_{0}^{\prime 2}}{a_{0}}\right) }{\left( a_{0}^{\prime 2}+a_{0}^{2}\right) 
\sqrt{1+\left( \frac{a_{0}^{\prime }}{a_{0}}\right) ^{2}}}.
\end{equation}%
We note that a prime stands for the differentiation with respect to $\theta
. $ It was found that with $\sigma _{0}\geq 0$ all energy conditions are
satisfied including that the matter which supports the wormhole was physical
i.e. not exotic. Finally the total matter contained in the throat can be
calculated as%
\begin{equation}
U=\int_{0}^{2\pi }a_{0}\sigma _{0}d\theta .
\end{equation}%
In the sequel we shall give explicit examples for this integral.

\begin{figure}[h]
\includegraphics[width=70mm,scale=0.7]{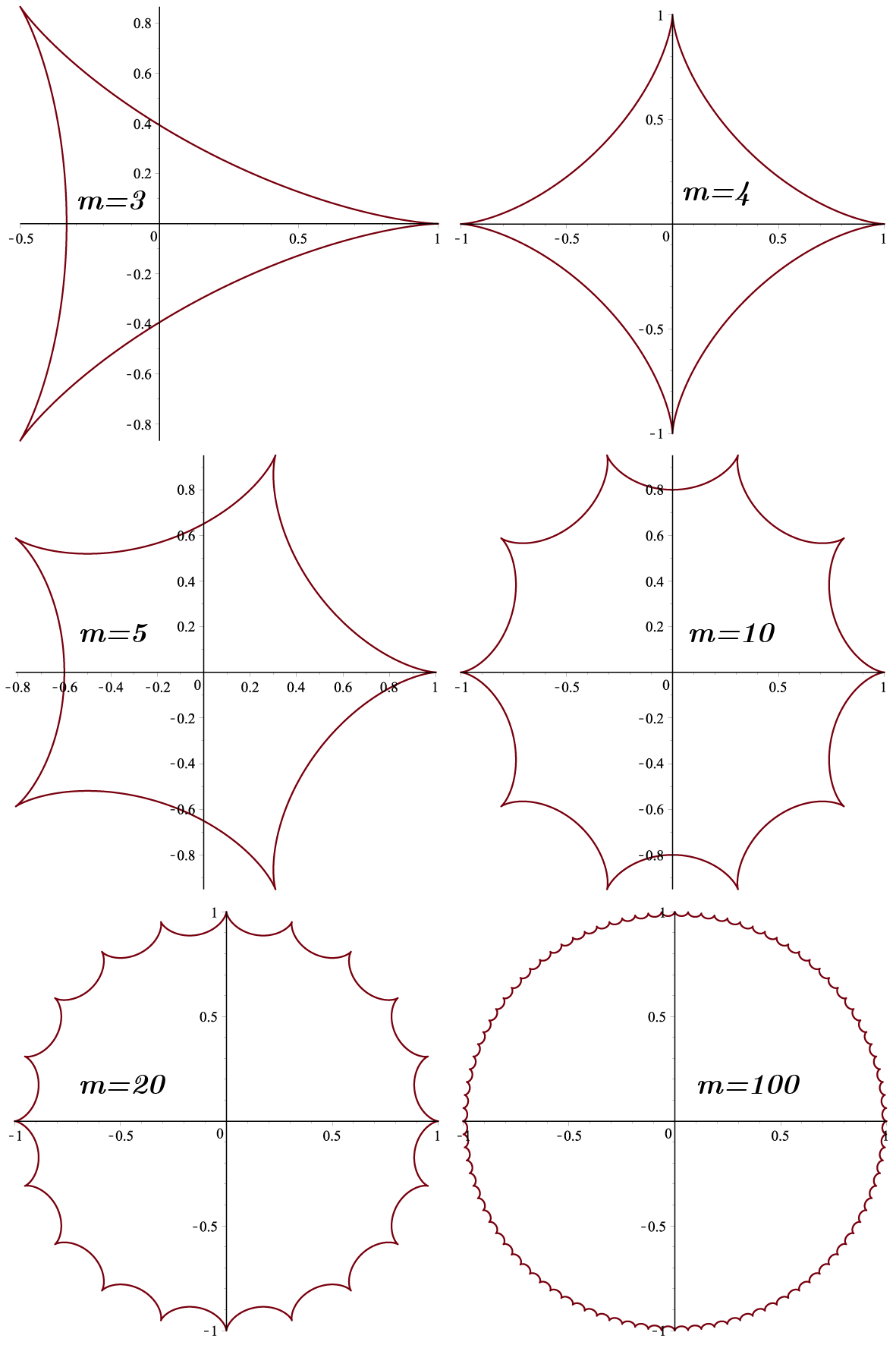}
\caption{Hypocycloid for different values of $m=3,4,5,10,20,100$.}
\end{figure}

\section{The Hypocycloid}

Hypocycloid \cite{Hyp} is the curve generated by a rolling small circle
inside a larger circle. This is a different version of the standard cycloid
which is generated by a circle rolling on a straight line. The parametric
equation of a hypocycloid is given by%
\begin{eqnarray}
x\left( \zeta \right)  &=&\left( B-b\right) \cos \zeta +b\cos \left( \frac{%
B-b}{b}\zeta \right)  \\
y\left( \zeta \right)  &=&\left( B-b\right) \sin \zeta -b\sin \left( \frac{%
B-b}{b}\zeta \right)   \notag
\end{eqnarray}%
in which $x$ and $y$ are the Cartesian coordinates on the hypocycloid. $B$
is the radius of the larger circle centered at the origin, $b\left(
<B\right) $ is the radius of the smaller circle and $\zeta \in \left[ 0,2\pi %
\right] $ is a real parameter. Here if one considers $B=mb,$ where $m\geq 3$
is a natural number, then the curve is closed and it possesses $m$
singularities / spikes. In Fig. 1 we plot (7) for different values of $m$
with $B=1.$ Let us add that for the particular choice of $B=1$ and $b=\frac{1%
}{4}$ the hypocycloid takes a compact form $x=\cos ^{3}\zeta $ and $y=\sin
^{3}\zeta $ with $x^{2/3}+y^{2/3}=1.$ In what follows we proceed to
determine the form of energy density $\sigma $ and the resulting total
energy for the individual cases plotted in Fig. 1.

To this end without loss of generality we set $B=1$ and $b=\frac{1}{m}$ and
express $\sigma $ as a function of $\zeta .$ For this we parametrize the
equation of the throat as 
\begin{eqnarray}
a &=&a\left( \zeta \right) =\sqrt{x\left( \zeta \right) ^{2}+y\left( \zeta
\right) ^{2}} \\
\theta &=&\theta \left( \zeta \right) =\tan ^{-1}\left( \frac{y\left( \zeta
\right) }{x\left( \zeta \right) }\right) .  \notag
\end{eqnarray}%
Using the chain rule one finds%
\begin{equation}
a^{\prime }=\frac{da}{d\theta }=\frac{\dot{a}}{\dot{\theta}}
\end{equation}%
and 
\begin{equation}
a^{\prime \prime }=\frac{d^{2}a}{d\theta ^{2}}=\frac{\ddot{a}\dot{\theta}-%
\dot{a}\ddot{\theta}}{\dot{\theta}^{3}}
\end{equation}%
which implies%
\begin{equation}
\sigma =\frac{1}{4\pi }\frac{a\ddot{a}\dot{\theta}-a\dot{a}\ddot{\theta}%
-a^{2}\dot{\theta}^{3}-2\dot{\theta}\dot{a}^{2}}{\left( \dot{a}^{2}+a^{2}%
\dot{\theta}^{2}\right) ^{\frac{3}{2}}}
\end{equation}%
where a dot stands for the derivative with respect to the parameter $\zeta .$
Consequently the total matter is given by%
\begin{equation}
U=\int_{0}^{2\pi }ud\zeta
\end{equation}%
where $u=a\sigma \dot{\theta}$ is the energy density per unit parameter $%
\zeta .$ Note that for the sake of simplicity we dropped the sub-index $0$
from the quantities calculated at the throat. Particular examples of
calculations for the energy $U$ are given as follows.

\subsection{$m=3$}

\begin{figure}[h]
\includegraphics[width=70mm,scale=0.7]{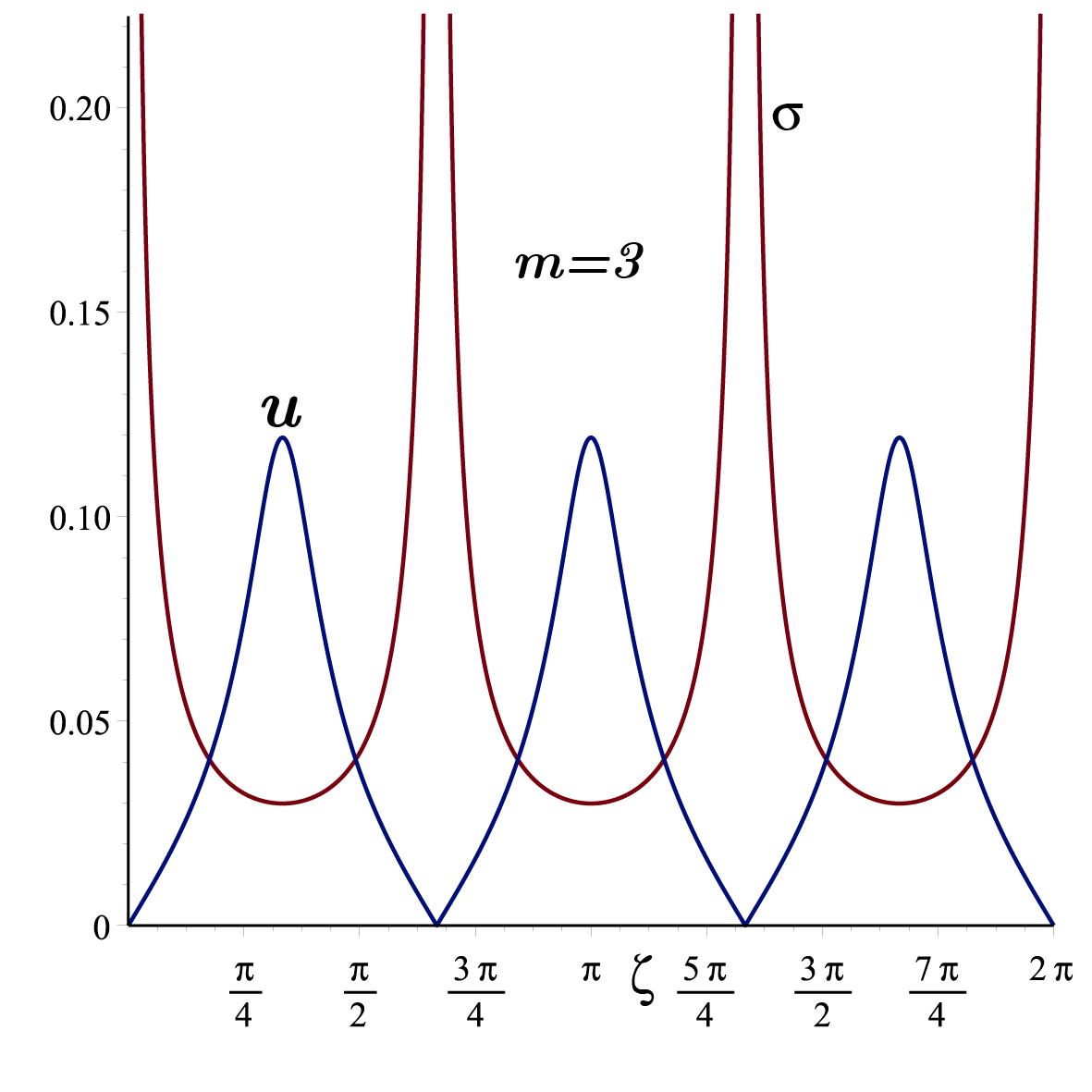}
\caption{Plot of $\protect\sigma $ and $u$ in terms of $\protect\zeta $ for $%
m=3.$ The singularities / cusps of $\protect\sigma $ are not physical. This
can be seen easily when the total energy is finite. This is in analogy with
a conical conductor of total charge finite but the charge density at the
vertex diverges. }
\end{figure}

The first case which we would like to study is the minimum index for $m$
which is $m=3.$ We find that%
\begin{equation}
\sigma =\frac{3\sqrt{2}}{32\pi \sqrt{\left( 1+2\cos \zeta \right) ^{2}\left(
1-\cos \zeta \right) }}
\end{equation}%
which is clearly positive everywhere. Knowing that the period of the curve
(7) is $2\pi $ we find that $\sigma $ is singular at the possible roots of
the denominator i.e., $\zeta =0,\frac{2\pi }{3},\frac{4\pi }{3},2\pi $. We
note that although $\sigma $ diverges at these points the function that must
be finite everywhere is $u$ which is given by%
\begin{equation}
u=\frac{3\sqrt{2}\sqrt{\left( 1+2\cos \zeta \right) ^{2}\left( 1-\cos \zeta
\right) }}{16\pi \sqrt{5-12\cos \zeta +16\cos ^{3}\zeta }}.
\end{equation}%
The situation is in analogy with the charge density of a charged conical
conductor whose charge density at the vertex of the cone diverges while the
total charge remains finite. In Fig. 2 we plot $\sigma $ and $u$ as a
function of $\zeta $ which clearly implies that $u$ is finite everywhere
leading to the total finite energy $U_{3}=0.099189.$

We would like to add that physically nothing extraordinary happens at the
cusp points. These points are the specific points at which the manifold is
not differentiable just with respect to $r$ but also with respect to the
angular variable $\theta .$ The original thin-shell wormhole has been
constructed based on discontinuity of the manifold with respect to $r$ at
the location of the throat which implied the presence of the matter source
at the throat (We refer to the Fig. 1 of Ref. \cite{MH3} where clearly such
cuspy point in $r$ direction is shown). Now, in the case under study we have
one additional discontinuity of the Riemann tensor in $\theta $ direction
which implies a more complicated form of matter distribution at the throat.

\subsection{$m=4$}

%
\begin{figure}[h]
\includegraphics[width=70mm,scale=0.7]{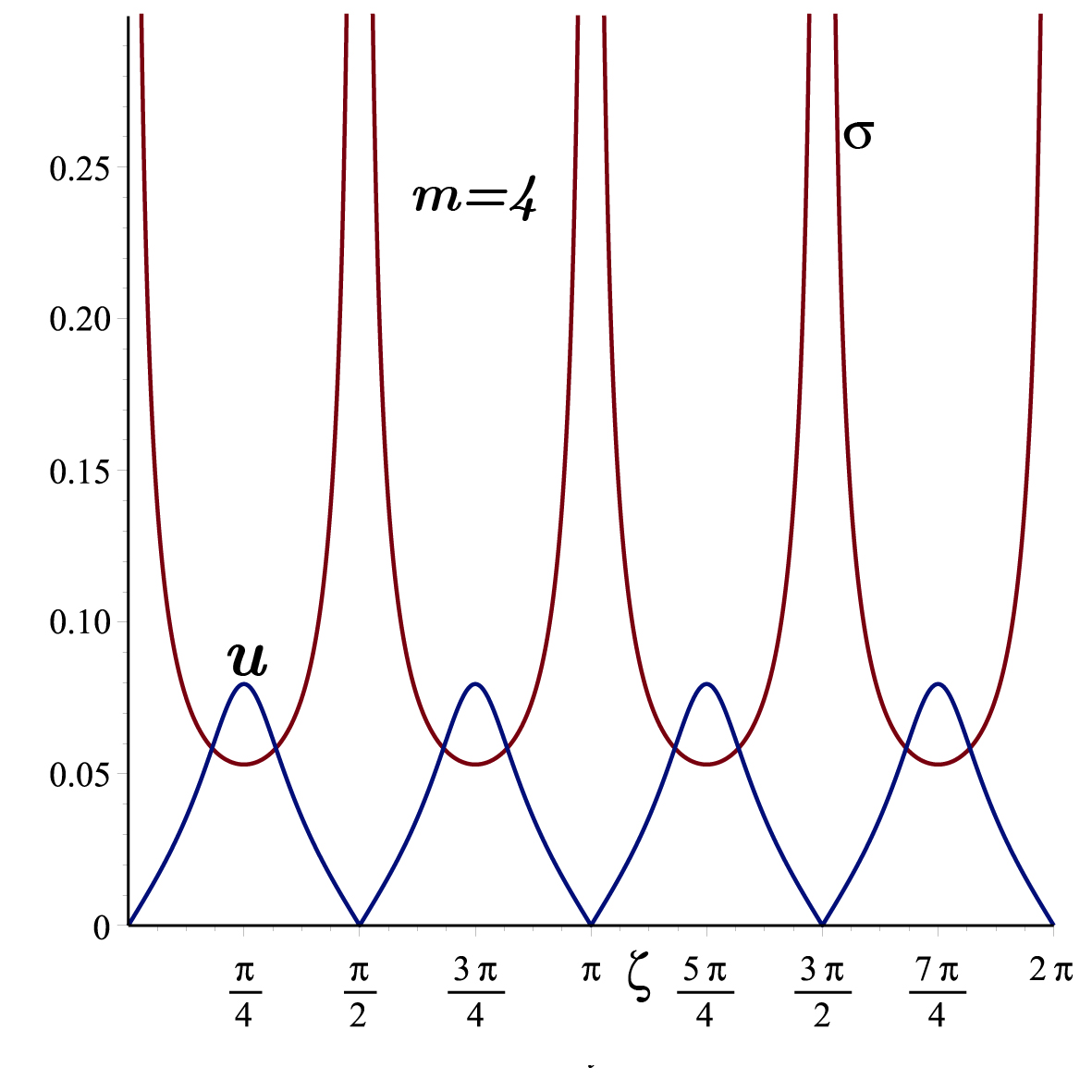}
\caption{Plot of $\protect\sigma $ and $u$ in terms of $\protect\zeta $ for $%
m=4.$ Similar to $m=3$ the singularities of $\protect\sigma $ are not
physical. }
\end{figure}

Next, we set $m=4$ where one finds%
\begin{equation}
\sigma =\frac{1}{6\pi \sqrt{\sin ^{2}\left( 2\zeta \right) }},
\end{equation}%
and%
\begin{equation}
u=\frac{\sqrt{\sin ^{2}\left( 2\zeta \right) }}{8\pi \sqrt{1-3\cos ^{2}\zeta
+3\cos ^{4}\zeta }}.
\end{equation}%
Fig. 3 depicts $\sigma $ and $u$ in terms of $\zeta $ and similar to $m=3$,
we find $\sigma >0$ and $u$ finite with the total energy given by $%
U_{4}=0.24203.$ As one observes $U_{4}>U_{3}$ which implies that adding more
cusps to the throat increases the energy needed. This is partly due to the
fact that the total length of the hypocycloid is increasing as $m$ increases
such that $\ell _{m}=\frac{8(m-1)}{m}$ with $B=1.$ This pattern goes on with 
$m$ larger and in general%
\begin{equation}
u=\frac{\left( m-2\right) ^{2}\sqrt{\left( \cos \zeta -\cos \left(
m-1\right) \zeta \right) ^{2}}}{8\pi \sqrt{2}\sqrt{\Psi }}
\end{equation}%
where 
\begin{multline}
\Psi =m^{2}-2\left( m-1\right) \cos ^{2}\left( m-1\right) \zeta \\
-\left( m-2\right) ^{2}\cos \zeta \cos \left( m-1\right) \zeta - \\
m^{2}\sin \left( m-1\right) \zeta \sin \zeta -2\left( m-1\right) \cos
^{2}\zeta .
\end{multline}

Table 1 shows the total energy $U_{m}$ for various $m.$ We observe that $%
U_{m}$ is not bounded from above (with respect to $m$) which means that for
large $m$ it diverges as $U_{m}\approx \frac{m}{2\pi }$. Therefore to stay
in classically finite energy region one must consider $m$ to be finite.

\begin{table}[th]
\caption{Total energy $U_{m}$ in terms of $m.$ We add that the total energy
for large $m$ is approximately given by $U_{m\rightarrow \text{large}%
}\approx \frac{m}{2\protect\pi }.$ This shows that increasing the number of
cusps to infinity requires infinite energy.}
\centering
\begin{tabular}{ccccccc}
\hline\hline
$m$ & $3$ & $4$ & $5$ & $10$ & $50$ & $100$ \\[1ex] \hline
$U_{m}$ & $0.099189$ & $0.24203$ & $0.39341$ & $1.1767$ & $7.5351$ & $15.492$
\\[1ex] \hline
\end{tabular}
\end{table}

\section{Conclusion}

The possibility of total positive energy has been scrutinized and verified
with explicit examples in the $2+1-$dimensional TSWs. Naturally the same
subject arises with more stringent conditions in the more realistic
dimensions of $3+1$. By getting advantage of technical simplicity we have
shown that the geometry of the throat can remarkably be that of a
hypocycloid. This is a rare curve compared with the more familiar minimum
time cycloid. In effect, a fixed point on the circumference of a smaller
circle rolling in a larger one makes the hypocycloid. The important point is
that in the rolling process concavity of the resulting curve makes the
extrinsic curvature negative, which in turn yields a positive energy density 
$\sigma .$ Note that with convex curves this is not possible. The emerging
cusps at the tips of the hypocycloid may yield singular points, however,
these can be overcome by integrating around such cusps. The lightning rod
analogy for diverging charge density in electromagnetism constitutes an
example to understand the situation. In the present case our sharp points
(edges) are reminiscent of cosmic strings and naturally deserves a separate
investigation. The results for the total energy turns out to be perfectly
positive, as our analytical calculation and numerical plots reveal.
Increasing frequency of each roll by using smaller and smaller circles
inside the large one is shown to increase the regular energy in finite
amounts which is necessary to give life to a TSW. The fact that in a static
frame the pressure vanishes simplifies our task. Here once $\sigma >0$ is
chosen it implies automatically that the energy conditions are also
satisfied. Finally, gluing together two curved spaces instead of flats will
be our next project to address in the same line of thought.

\appendix

\section{Extrinsic curvature tensor}

The bulk metric is flat given by (1), therefore we cut out $r<a\left( \theta
\right) $ from the bulk and make two identical copies of the rest manifold.
We paste them at the timelike hypersurface $F\left( r,\theta \right)
=r-a\left( \theta \right) =0$ to construct a complete manifold. The induced
metric on the hyperplane $\Sigma $ is given by (3). The extrinsic curvature
tensor on the shell $\Sigma $ is given by%
\begin{equation}
K_{ij}^{\left( \pm \right) }=-n_{\gamma }^{\left( \pm \right) }\left( \frac{%
\partial ^{2}x^{\gamma }}{\partial y^{i}\partial y^{j}}+\Gamma _{\alpha
\beta }^{\gamma }\frac{\partial x^{\alpha }}{\partial y^{i}}\frac{\partial
x^{\beta }}{\partial y^{j}}\right)
\end{equation}%
in which $x^{\gamma }=\left( t,r,\theta \right) $ is the coordinate of the
bulk metric and $y^{i}=\left( t,\theta \right) $ is the coordinate of the
shell. Also%
\begin{equation}
n_{\gamma }^{\left( \pm \right) }=\frac{\pm 1}{\sqrt{\Delta }}\frac{\partial
F}{\partial x^{\gamma }}
\end{equation}%
where%
\begin{equation}
\Delta =g^{\alpha \beta }\frac{\partial F}{\partial x^{\alpha }}\frac{%
\partial F}{\partial x^{\beta }}
\end{equation}%
refers to the normal $3-$vector to the shell and $\pm $ implies the
different sides of the shell.

The Israel junction \cite{Israel} conditions read%
\begin{equation}
-8\pi S_{i}^{j}=k_{i}^{j}-\delta _{i}^{j}k
\end{equation}%
in which $S_{i}^{j}=diag\left( -\sigma ,p\right) $ is the energy-momentum
tensor on the shell (we note that the off diagonal term is zero) and $%
k_{i}^{j}=$ $K_{i}^{j\left( +\right) }-K_{i}^{j\left( -\right) }$ with $%
k=k_{i}^{i}.$ The explicit calculation reveals that%
\begin{equation}
n_{\gamma }^{\left( \pm \right) }=\pm \frac{a}{\sqrt{a^{2}+a^{\prime 2}}}%
\left( 0,1,-a^{\prime }\right)
\end{equation}%
and%
\begin{equation}
k_{i}^{j}=\left[ 
\begin{array}{cc}
0 & 0 \\ 
0 & \frac{2\left( a^{2}+2a^{\prime 2}-aa^{\prime \prime }\right) }{\left(
a^{2}+a^{\prime 2}\right) ^{3/2}}%
\end{array}%
\right] .
\end{equation}

\bigskip

\end{document}